\newcommand*{\ATLASLATEXPATH}{latex/}
\documentclass[cernpreprint,texlive=2011,txfonts,UKenglish]{latex/atlasdoc}

\usepackage{\ATLASLATEXPATH atlaspackage}

\usepackage{longtable}
\usepackage{multirow}
\usepackage{lscape}
\usepackage{verbatim}
\usepackage{float}
\usepackage{graphicx}
\usepackage{url}
\usepackage{caption}
\usepackage{hyperref}
\newsavebox{\tempbox}
\usepackage{epstopdf}
\usepackage{booktabs}

\usepackage{geometry}
\usepackage{changepage}
\usepackage{numprint}
\usepackage{longtable}
\usepackage{array}
\usepackage{color}
\makeatletter
\usepackage{epstopdf}
\usepackage{\ATLASLATEXPATH atlasbiblatex}

\usepackage{\ATLASLATEXPATH atlascontribute}

\usepackage{\ATLASLATEXPATH atlasphysics}

\graphicspath{{logos/}{figures/}}

\usepackage{ScalarLQPaper-defs}

\AtlasTitle{Search for scalar leptoquarks in $pp$ collisions at \\$\boldmath{\sqrt{s}}$ = 13~TeV with the ATLAS experiment}

\author{The ATLAS Collaboration}

\AtlasRefCode{EXOT-2015-19}

\PreprintIdNumber{CERN-EP-2016-074}

\AtlasJournal{New Journal of Physics}
\AtlasJournalRef{New J. Phys. 18 (2016) 093016}
\AtlasDOI{10.1088/1367-2630/18/9/093016}

\AtlasAbstract{%
An inclusive search for a new-physics signature of lepton-jet resonances has been performed by the ATLAS experiment. Scalar
leptoquarks, pair-produced in $pp$ collisions at  $\sqrt{s}$ = 13~TeV  at the Large Hadron Collider, have been considered. An  integrated luminosity of 3.2\,\ifb,
corresponding to the full 2015 dataset was used. First (second) generation leptoquarks were sought in events with two electrons (muons) and two or more jets. The
observed event yield in each channel is consistent with Standard Model background expectations. The observed (expected) lower limits on the leptoquark mass at 95\% confidence level are 1100 GeV and 1050 GeV (1160 GeV and 1040 GeV) for first and second generation leptoquarks, respectively, assuming a branching ratio into a
charged lepton and a quark of 100\%. Upper limits on the aforementioned branching ratio are also given as a function of leptoquark mass. Compared with the results of earlier ATLAS searches, the sensitivity is increased for leptoquark masses above 860 GeV,  and the observed exclusion limits confirm and extend the published results.
}

\hypersetup{pdftitle={ATLAS document},pdfauthor={The ATLAS Collaboration}}

\begin{document}

\maketitle

\tableofcontents


\newpage
\section{Introduction}\label{sec:intro}

The Large Hadron Collider (LHC) Run 2 has provided the possibility to study $pp$ collisions at 13~TeV centre-of-mass energy for the first time, and has thus opened a new
discovery window for physics beyond the Standard Model (SM). The presented analysis is an inclusive search for new physics phenomena resulting in final state signatures of lepton-jet resonances in
the first 3.2\,fb$^{-1}$ of 13~TeV data collected by the ATLAS detector. Such
phenomena may not
have been kinematically accessible at the lower Run 1 centre-of-mass energy of 8 TeV.
As a benchmark signal model, scalar leptoquarks decaying to jets and leptons were used.

 Leptoquarks (LQ) feature in a number of  theories~\cite{BSM,techni2,techni3,string,comp,pati_salam_colour,georgi_glashow_unification} which extend the SM, such as grand unified theories and models with quark and lepton substructure.  LQs possess non-zero baryon and lepton numbers and their existence would provide a connection between quarks and leptons. This could help explain the observed similarity of the quark and lepton sectors in the SM. LQs carry a colour-triplet charge and a fractional electric charge~\cite{Schrempp:1984nj}.
 They can be scalar or vector bosons and they decay directly to lepton--quark pairs. The analysis presented in this paper focuses on the pair production of scalar leptoquarks.
 \begin{figure}[h]
  \centering
  \includegraphics[width=0.3\textwidth]{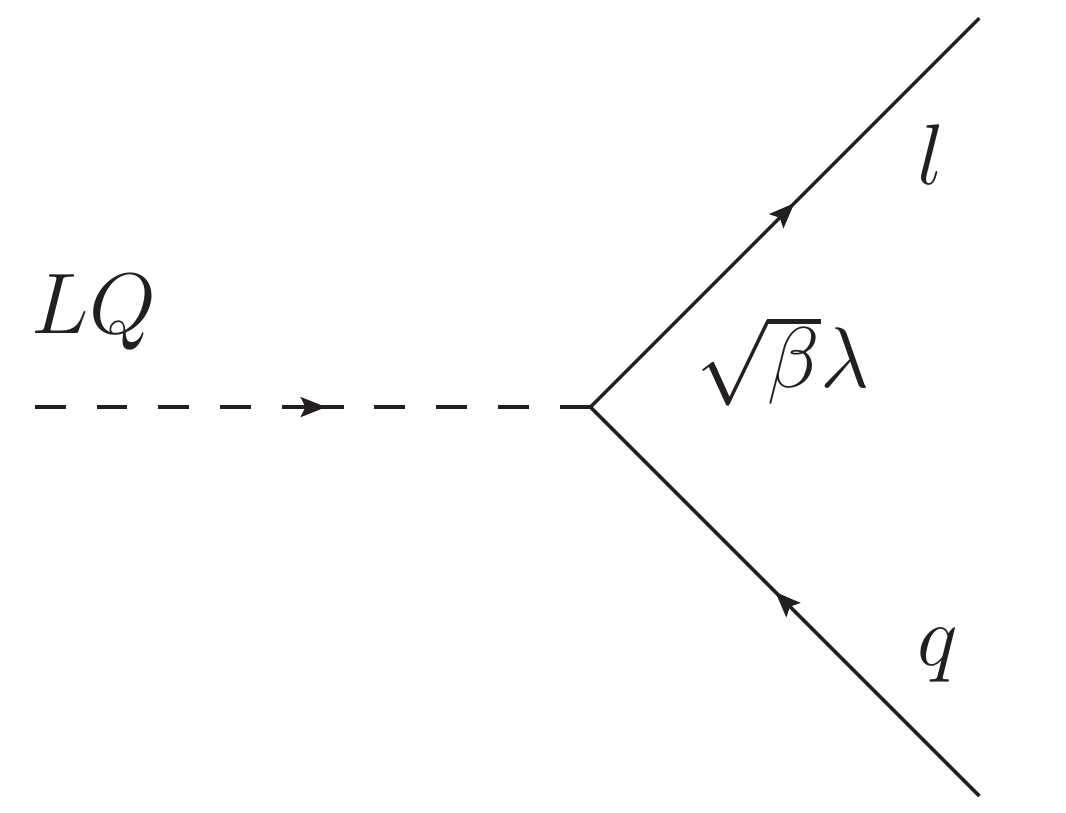}
  \caption{Feynman diagram showing the Yukawa coupling $\lambda_{\ell}=\sqrt{\beta}\lambda$ between a leptoquark, a lepton ($\ell$) and a quark ($q$).}
  \label{fig:lamdaLQ}
\end{figure}

\begin{figure}[h]
  \centering
  \includegraphics[width=0.6\textwidth]{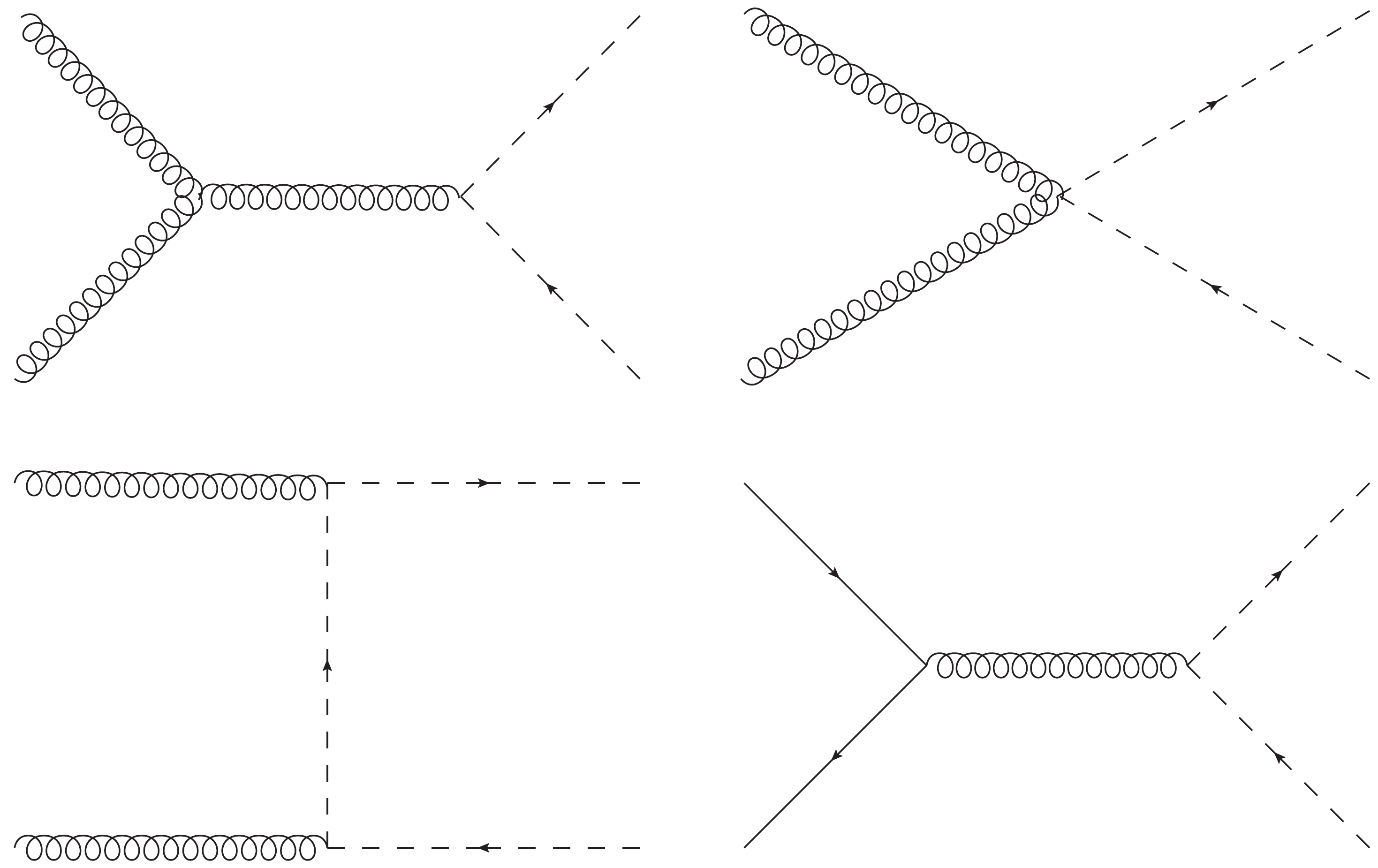}
   \caption{Dominant leading-order Feynman diagrams for the pair production of scalar leptoquarks from gluon fusion and quark--antiquark annihilation.}
  \label{fig:FeynLQ}
\end{figure}

A single Yukawa coupling ($\lambda_{\ell}$) governs the interaction strength between a scalar LQ and a given quark ($q$) and lepton ($\ell$) pair. 
A Feynman diagram showing a LQ decay is shown in Figure~\ref{fig:lamdaLQ}. 
The couplings are determined by two free parameters of the model: the branching ratio into charged leptons, $\beta$, and the coupling parameter, $\lambda$. 
The coupling to a charged lepton and a quark is given by $\lambda_\ell=\sqrt{\beta}\lambda$, the coupling to a neutrino and a quark by $\lambda_\nu=\sqrt{1-\beta}\lambda$.
The
pair-production cross section of leptoquarks in \textit{pp} collisions is largely insensitive to the coupling values, since the basic processes of $\LQ$ pair-production are
gluon fusion and quark--antiquark annihilation. Example leading-order diagrams are shown in Figure~\ref{fig:FeynLQ}. At a centre-of-mass energy of $\sqrt{s} = 13$\,\TeV,
gluon fusion is the dominant process. For $\LQ$ masses (\mlq) up to a few hundred GeV, it contributes up to 95\% of the total cross section. Above
\mlq$=1.5$~TeV, the contribution from quark--antiquark annihilation amounts to about 30\%~\cite{Kramer:2004df}. 
Therefore, the parameter of interest -- apart from the $\LQ$ mass -- is the branching ratio $\beta$.

The signal benchmark model for LQ production used in this analysis is the minimal
Buchm\"uller--R\"uckl--Wyler
model (mBRW)~\cite{Buchmuller:1986zs}. In this approach a number of constraints are imposed on
 the LQ properties. Lepton number and baryon number are separately conserved to prevent fast proton decay. The LQ couplings are also considered to be purely chiral. Furthermore, it is assumed that LQs belong to three generations (first, second and third) which interact only with lepton--quark pairs within the same generation. With this assumption, lepton-flavour violation
 is suppressed. 
 However, in a more generic picture of leptoquarks, a LQ may couple to a quark and a lepton belonging to different generations~\cite{Agashe:2014kda}. 
 Although the results of this search were not explicitly interpreted in this type of model, the event selections used were designed to retain sensitivity to leptoquark models in which decays into first or second generation leptons and bottom-quarks ($b$) are possible.

Previous searches for pair-produced LQs have been performed by the ATLAS and CMS
collaborations~\cite{CMS-EXO-10-005,CMS-EXO-10-007,EXOT-2010-08,EXOT-2011-17,EXOT-2011-26,CMS-EXO-11-028,CMS-EXO-12-002,EXOT-2012-14,EXOT-2014-03,CMS-EXO-12-041,CMS-EXO-11-030,CMS-EXO-12-032,CMS-EXO-14-008}
 at $\sqrt{s}$~=~7 and 8~TeV. The existence of scalar LQs with masses up to 1050~GeV and 1000~GeV (for $\beta=1$)
 for first-  and second-generation scalar LQs, respectively, is excluded at 95\% confidence level (CL) by ATLAS~\cite{EXOT-2014-03} in a
 study performed at $\sqrt{s}$~=8~TeV using 20~fb$^{-1}$ of integrated luminosity.
 The CMS experiment similarly excluded first- and second-generation scalar leptoquarks  up to masses of
 1010~GeV and 1080~GeV (for $\beta=1$), respectively~\cite{CMS-EXO-12-041}.

In this paper, searches for the pair-production of
leptoquarks of the first (\LQone) and second (\LQtwo) generations,
based on events containing exactly two
electrons or muons and at least two jets
(denoted by \eejj~and \mumujj, respectively), are reported. In order
to keep the
search as inclusive as possible, it was not required that the charges of the
two leptons in an event must be opposite. Similarly, no selections on the jet flavour were introduced so as not to exclude possible $\LQ\rightarrow eb$ and $\LQ\rightarrow\mu b$ decays.

\section{The ATLAS detector}

The ATLAS experiment \cite{PERF-2007-01} is a multi-purpose  detector with a forward-backward symmetric cylindrical geometry and nearly 4$\pi$ coverage in solid  angle\footnote{ATLAS uses a right-handed coordinate system with its origin at the nominal interaction point (IP) in the centre of the detector and the $z$-axis along the beam pipe. The $x$-axis points from the IP to the centre of the LHC ring, and the $y$-axis points upward. Cylindrical coordinates $(r,\phi)$ are used in the transverse plane, $\phi$ being the azimuthal angle around the $z$-axis. The pseudorapidity is defined in terms of the polar angle $\theta$ as $\eta=-\ln\tan(\theta/2)$.}.  The three major sub-components of ATLAS are the tracking detector, the calorimeter and the muon spectrometer.  Charged-particle tracks and vertices are reconstructed by the inner detector (ID) tracking system, comprising silicon pixel (including the newly installed innermost pixel layer), and microstrip detectors covering the pseudorapidity range $|\eta|$~$<$~2.5, and a straw tube tracker that covers $|\eta|$~$<$~2.0.  The ID is immersed in a homogeneous 2\,T magnetic field provided by a solenoid.  Electron, photon, jet and $\tau$ lepton energies are measured with sampling calorimeters.  The ATLAS calorimeter system covers a pseudorapidity range of $|\eta|$~$<$~4.9.  Within the region $|\eta|$~$<$~3.2, electromagnetic calorimetry is provided by barrel and endcap high-granularity lead/liquid argon (LAr) calorimeters, with an additional thin LAr presampler covering $|\eta|$~$<$~1.8, to correct for energy loss in material upstream of the calorimeters. Hadronic calorimetry is provided by a steel/scintillator-tile calorimeter, segmented into three barrel structures within $|\eta|$~$<$~1.7, and two copper/LAr hadronic endcap calorimeters. The forward region (3.1~$<$~$|\eta|$~$<$~4.9) is instrumented by a LAr calorimeter with copper (electromagnetic) and tungsten (hadronic) absorbers.  Surrounding the calorimeters is a muon spectrometer (MS) with superconducting air-core toroids, providing bending powers of 3\,Tm in the barrel and 6\,Tm in the endcaps. The MS includes a system of precision tracking chambers providing coverage over $|\eta|$~$<$~2.7. Three stations of precision tracking chambers are used to measure the curvature of tracks. The MS also contains detectors with triggering capabilities over $|\eta|$~$<$~2.4 to provide fast muon identification and momentum measurements.

The ATLAS two-level trigger system is used to select events considered in this paper.
The first-level trigger is hardware-based while the second, high-level trigger is implemented in software and employs algorithms similar to those used offline in the full event reconstruction.

\section{Signal and background simulations}
\label{sec:DataMC}

The \pythia 8.160~\cite{Sjostrand:2007gs} Monte Carlo (MC) model, based on leading-order (LO) matrix-element calculations supplemented with parton showers, was used with the ATLAS{\sc A14}~\cite{ATL-PHYS-PUB-2014-021} set of tuned parameters (tune) for the underlying event, together with the {\sc NNPDF23LO}~\cite{Ball:2012cx}  parton distribution  functions (PDFs), to produce simulated samples of pair-produced first- and second-generation scalar LQs.  Leptoquarks of the first (second) generation decay to $e^{+}e^{-}u\bar{u}$ ($\mu^{+}\mu^{-}c\bar{c}$)  final states. 
 Samples were produced for $\LQ$ masses in the
range of 500--1500~GeV. As was also done in the previous ATLAS publication~\cite{EXOT-2014-03}, the value of the coupling parameter $\lambda$ was set to
$\sqrt{0.01\times4\pi\alpha}$, where $\alpha$ is the fine-structure
constant. This value of $\lambda$ determines the leptoquark natural width, which is less than 100\,\MeV~and is smaller than the detector resolution for the reconstruction of leptoquark mass. It also leads to a $\LQ$ lifetime sufficiently small such that $\LQ$s in the mass range considered in this work would decay promptly.  Next-to-leading-order (NLO) calculations~\cite{Kramer:2004df} of the cross sections for scalar leptoquark pair-production were used to normalise the signal samples.

The dominant SM backgrounds arise from processes which can produce a final state containing two reconstructed high transverse
momentum ($\pt$) leptons (electrons or muons) and jets. Simulated samples were made of Drell--Yan production ($q\bar{q} \rightarrow Z/\gamma^* \rightarrow {\ell}^+{\ell}^-$) and the production of  $t\bar{t}$,  diboson ($WW$, $WZ$, and $ZZ$) and single top-quarks in association with a $W$ boson.

Drell--Yan events with associated jets were simulated using the {\sc Sherpa} 2.1.1~\cite{Gleisberg:2008ta} generator.
Matrix elements were calculated for up to two partons at NLO and four partons at LO using the {\sc Comix}~\cite{Gleisberg:2008fv} and {\sc OpenLoops}~\cite{Cascioli:2011va} matrix-element generators and merged with the {\sc Sherpa} parton shower~\cite{Schumann:2007mg} using the
 {\sc ME+PS@NLO} prescription~\cite{Hoeche:2012yf}.
The {\sc CT10} PDF set~\cite{Lai:2010vv} was used in conjunction with dedicated parton-shower tuning developed by the authors of {\sc Sherpa}~\cite{Schumann:2007mg}.

For the generation of \ttbar{} and single top quarks in the $Wt$ channel, the {\sc Powheg-Box} v2 generator~\cite{Nason:2004rx,Frixione:2007vw,Alioli:2010xd,Re:2010bp} with the {\sc CT10} PDF set in the matrix-element calculations was used.
For both processes the parton shower, fragmentation, and the underlying event were simulated using \pythia~6.428~\cite{bib:Pythia6} with the {\sc Perugia} 2012 tune~\cite{Skands:2010ak} and using
the {\sc Cteq6L1} PDF set~\cite{Pumplin:2002vw}.
The top-quark mass was set to 172.5~GeV.
The {\sc EvtGen} v1.2.0 program~\cite{EvtGen} was used to simulate the bottom and charm hadron decays.

Diboson processes with four charged leptons, three charged leptons + one neutrino or two charged leptons and two neutrinos were simulated using the {\sc Sherpa} 2.1.1 generator. Matrix
elements contain all diagrams with four electroweak vertices. They were calculated for up to one ($4\ell$, $2\ell+2\nu$) or zero partons ($3\ell+1\nu$) at NLO and up to three partons at LO using the {\sc Comix} and {\sc OpenLoops} matrix element generators and merged with the {\sc Sherpa} parton shower using the ME+PS@NLO prescription. Diboson processes with one of the bosons decaying hadronically and the other leptonically were simulated using the same {\sc Sherpa} version.

All samples of simulated events include the effect of multiple proton--proton interactions in the same or neighbouring bunch crossings (pile-up) which were modelled by overlaying simulated minimum-bias events on each generated signal and background event.
These multiple interactions were simulated with the soft QCD processes of \pythia 8.186~\cite{Sjostrand:2007gs} using tune {\sc A2}~\cite{ATL-PHYS-PUB-2012-003} and the
{\sc MSTW2008LO} PDF set~\cite{mstw}.
The number of overlaid events was chosen to match the average number of interactions per $pp$ bunch crossing observed in the data as it evolved throughout the data-taking period (giving an average of 14 interactions per crossing for the whole data-taking period).
The SM background samples were processed through the  {\sc GEANT4}-based detector simulation~\cite{SOFT-2010-01,Geant4}, while a fast simulation using a parameterisation of the performance of the calorimeters~\cite{ATL-PHYS-PUB-2010-013} and {\sc GEANT4} for the other parts of the detector was used for the signal samples and some samples used for studies of systematic uncertainties. The standard ATLAS reconstruction software was used for both simulated and collision data.

 Estimates of the cross sections of background processes were taken from the following theoretical predictions.
Single-top
production was calculated at NLO+next-to-next-to-leading-logarithm (NNLL) accuracy~\cite{MCatNLO_Wt}.
 Estimates of Drell--Yan and $t\bar{t}$ production cross sections at NLO~\cite{Gleisberg:2008ta} and NLO+NNLO~\cite{Czakon:2011xx} accuracy, respectively, were used.

\section{Physics object definition}
\label{sec:ObjSel}
The electron energy was measured using its associated cluster of electromagnetic-calorimeter cells with significant energy deposits,  whereas the direction was determined by the track associated with this cluster.
To identify and select electrons, requirements were placed on the shape of the cluster, on the quality of the associated track, and on the degree
of matching between the track and cluster.  Electron candidates must have transverse energy \ET{} $>30$\,\GeV~and $|\eta|<2.47$.  Electron candidates associated with clusters in the transition region between the barrel and endcap calorimeters ($1.37$~<~$|\eta|$~<~$1.52$) were not considered.  All electrons
must be reconstructed with a cluster-based or a combined cluster- and track-based algorithm~\cite{ATLAS-CONF-2014-032}. Furthermore, the impact parameters of the electron track relative
to the beam line were required to satisfy  $|d_0 / \sigma_{d_0}| < 5$ and $|z_0\sin \theta| < 0.5$~mm, where $d_0$, $\sigma_{d_0}$ and $z_0$ are the transverse impact parameter, its
uncertainty, and the longitudinal impact parameter,
respectively. In addition, electron isolation
requirements were imposed on the summed transverse momentum of tracks (transverse energy of clusters) in a cone around the electron track (cluster barycentre). 
The radius of the cone around the track is $\Delta R = 10~\mathrm{GeV} / p_\mathrm{T}$ for $p_\mathrm{T} > 50 \,\mathrm{GeV}$ and 0.2 otherwise.\footnote{ Here, $\Delta R =
\sqrt{(\Delta\eta)^2 + (\Delta\phi)^2}$ is a cone defined by differences in pseudorapidity and azimuthal angle. $p_{\rm T}$ is given in GeV.}
For the cluster isolation, a fixed cone radius size of 0.2 is used.
The efficiency of these isolation criteria is higher than 99\%.
The reconstruction efficiency is higher
than 98\% in most regions of transverse momentum and pseudorapidity. The identification efficiency varies between 75\% and 92\%, rising as a function of \et{}~\cite{el_id_2015}. All of these efficiencies refer to the efficiency for a single electron, independent of the specific event topology. 

Muon tracks were reconstructed independently in the ID and the MS.
These muon tracks were required to have a minimum number of associated hits in each system and to satisfy geometrical and momentum matching criteria.
The two tracks were then used as input to a combined fit which takes into account the energy loss in the calorimeter and multiple-scattering effects~\cite{PERF-2014-05}.
To improve momentum resolution and ensure a reliable measurement at very high momenta,
muon tracks were required to have at least three hits in each of the three precision chambers in the MS.
Tracks which traverse precision chambers with poor alignment were rejected.
Finally, measurements of charge over momentum, performed independently in the ID and MS, were required to agree within seven standard deviations of the sum in quadrature of the uncertainties in the corresponding ID and MS measurements.
Muon candidates were required to have \pt~$>$~40~GeV and
$|\eta|$~<~$2.5$. Those falling in the overlap region of the MS barrel and endcap ($1.01<|\eta|<1.10$) were rejected due to the potential \pt\ mismeasurement
resulting from relative barrel--endcap misalignment.
Muon candidates were required to fulfil $|d_0 / \sigma_{d_0}|$~<~$3$ and $|z_0\sin \theta|$~<~$0.5$~mm. In order to reduce the background from light- and heavy-hadron decays within jets,
muons were required to fulfil isolation requirements. The track-based isolation variable used is the sum of the transverse
momenta of the tracks in a cone around the muon of size $\Delta R= 10$~GeV/$p_{\rm T}^\mu$, excluding the muon itself and with $p_{\rm T}^\mu$ in GeV.
The isolation efficiency is greater than 99\%.
The efficiency for reconstructing and identifying muons using the criteria described above is typically
greater than 80\% for the transverse momentum and pseudorapidity selections used in this work~\cite{PERF-2015-10}. As for electrons, these efficiencies refer to single objects.

The anti-$k_{t}$ algorithm~\cite{AntiKt} with a radius parameter $R=0.4$  was used to reconstruct jets from energy clusters in the
calorimeter~\cite{PERF-2014-07}. Jet calibration is performed using energy- and $\eta$-dependent correction factors derived from simulations
together with further corrections from {\it in situ} measurements.
The jets used in this work must satisfy $\pt$~>~$50$~GeV and $|\eta|$~<~$2.8$. Further selections were applied to ensure that all jets considered are well measured~\cite{ATLAS-CONF-2015-029}. A description of the jet energy scale measurement and its associated systematic uncertainties can be found in Ref.~\cite{PERF-2012-01}.
Jet-flavour tagging techniques were not used in this paper and, as a consequence, good sensitivity was maintained for LQ decays into a lepton plus any quark flavour barring the top quark.

Ambiguities in the object identification during reconstruction, i.e. when a reconstructed object matched multiple object identification hypotheses  (electron, muon, jet), were resolved in the following way.
First, electrons were removed if they shared their track with a muon. In a second step, ambiguities between electrons
and jets were removed; if the two objects had $\Delta R < 0.2$ the jet was rejected; if $0.2 < \Delta R < 0.4$ the electron
was rejected. Finally, muon--jet ambiguities were resolved as follows: if the muon and jet were closer than $\Delta R = 0.4$
the jet was rejected if it has less than three tracks, otherwise the muon was rejected.
For the definition of signal and control regions (see Section~\ref{sec:Strategy}) only objects remaining after this procedure were considered.

\section{Dataset and event selection}
\label{sec:EvtSel}

Proton--proton collision data at a centre-of-mass energy of \sqrts~=~13\,\TeV, collected by the ATLAS detector at the LHC during 2015, were used. After applying data quality criteria, the dataset corresponds to an integrated luminosity of 3.2\,\ifb.

Events considered in the search were selected by the ATLAS two-level trigger system~\cite{ATL-DAQ-PUB-2016-001}.
In the \eejj channel, a two-electron trigger was used with an \et\ threshold of 17\,\GeV{} for each electron.
The \mumujj  search used events selected by either of two single-muon triggers.
The first trigger has a muon \pt{} threshold of 26\,\GeV{} and additional requirements on its properties.
In particular, it requires the muon to be isolated, which leads to a loss in efficiency at high \pt{}.
To retain a high trigger efficiency in the region of high \pt{}, the second trigger,
which has a  \pt{} threshold of 50\,\GeV{} but no additional requirements, was used. The trigger efficiencies for the \eejj and  \mumujj searches  exceed 90\% for the object kinematics considered in this analysis.

Multiple $pp$ interactions during bunch crossings lead to events containing a number of reconstructed vertices. The primary vertex of the event 
is defined as that vertex with the largest sum of squared transverse momenta of its associated tracks. Events which contain a primary vertex with at least two associated tracks satisfying $p_{\mathrm{T,track}} >$ 0.4 GeV were selected. Furthermore, MC events were given a per-event weight to correct for differences in the distribution of the average number of $pp$ interactions per bunch crossing between data and simulation.

Only events with exactly two charged leptons and at least two jets were considered for this analysis.
Scale factors were applied as event weights to correct the MC description of lepton trigger, reconstruction, identification, isolation and impact-parameter cut efficiencies.
A description of the derivation of the scale factors, obtained by comparing data and MC predictions in dedicated studies, can be found in Refs.~\cite{ATL-PHYS-PUB-2015-037,ATLAS-CONF-2014-032}.

\section{Analysis strategy: signal, control and validation regions}\label{sec:Strategy}
The analysis presented here used signal (SR), control (CR) and validation (VR)  regions to
optimise
signal significance and to constrain the normalisation of the main background sources.
The latter are Drell--Yan events containing $Z/\gamma^* \rightarrow {e^+}{e^-},{\mu^+}{\mu^-}$+jets processes (hereafter termed DY+jets) and $t\bar{t}$ events in which both top quarks decay leptonically.
The signal, control and validation regions were defined using the following discriminating observables:
\begin{itemize}
\item  The dilepton invariant mass: $m_{\ell \ell}$.
\item  The scalar sum \st{} of the transverse momentum of the two leptons and of the two leading jets.
\item  The minimum invariant mass of the two lepton--jet pairs in an event, $\mlqmin$.  The lepton--jet pairs were chosen such that the invariant mass difference between them was smallest. The lower mass of the two combinations
was chosen as the discriminating variable following dedicated sensitivity studies.
\end{itemize}

The signal region was defined by requiring $m_{\ell \ell}>130$ GeV and \st{}~$>$~600 GeV.
The cut on $m_{\ell \ell}$ was chosen to reduce the DY+jets background.
The cut on \st{} was optimised by maximising the discovery significance~\cite{Cowan:2010js} for LQs with masses between $500$ GeV and $1500$ GeV, i.e. by performing a likelihood fit (described in more detail at the end of this section) using the {\mlqmin}~distribution defined above for \st{} values between $0$ and $3$~TeV in steps of $100$~GeV. This study showed that there is little dependence of the optimised \st{} value on the LQ mass when using the shape information of the mass spectrum. It was confirmed that the approach used in previous results, i.e. a cut-and-count analysis in several signal regions defined by varying cuts on the three variables mentioned above, does not give better sensitivity.

The signal selection efficiency is defined as the fraction of all simulated signal events, generated across the full phase space, that survive the trigger and the
final SR selection.  For leptoquarks of the first (second) generation, the overall selection efficiency rises from around 62\%  to 71\% (38\% to 43\%) as the mass increases from 500~GeV to 1500~GeV. The lower efficiency for second-generation leptoquarks is due to the muon track requirements, which demand hits in three MS stations, and which are needed to give an optimal momentum resolution at high $p_T$ for this analysis.

Three non-overlapping control regions with negligible signal contamination
were defined.
Differences in the predicted and observed event yields in these regions were
used to evaluate scale factors which were used to normalise the MC predictions for the  $t\bar{t}$ and DY+jets backgrounds in the SRs. 
The two DY+jets CRs -- one for the \eejj and one for the \mumujj channel --  were defined by requiring at least two jets and exactly two same-flavour leptons with a dilepton invariant mass restricted to a window around the $Z$ boson mass: $70<m_{\ell \ell}<110$~GeV.
The $t\bar{t}$ control region requires at least two jets, exactly one muon and exactly one electron: these events were selected with the same single-muon triggers as described in Section~\ref{sec:EvtSel}.
This control region is common to both channels.

Validation regions were used to verify that data and MC predictions agree in a phase space close to the signal regions, but still with a negligible signal contamination. This was achieved by applying the same selection as for the signal region, but inverting the cut on \st{}, i.e. allowing only values below 600~GeV.

The requirements for the various regions are collected in Table~\ref{tab:regionDefs}.
\begin{table}
  \centering
  \caption{Definition of control, signal and validation regions. In all regions, at least two jets were required.}
  \begin{tabular}{|c|c|c|c|c|c|}
    \hline\hline
    Region & Channel & \#e & \#$\mu$ & \mll{}\,[GeV] & \st{}\,[GeV] \\
    \hline
    \ttbar{} CR & both & 1 & 1 &--   & -- \\ \hline
    \multirow{2}{*}{DY+jets CR} & $eejj$ & 2 & 0 & \multirow{2}{*}{$[70, 110]$}   & \multirow{2}{*}{--}  \\
                                   & $\mu\mu jj$ & 0 & 2 &    &  \\ \hline
    \multirow{2}{*}{SR} & $eejj$ & 2 & 0 & \multirow{2}{*}{$>$130}   &  \multirow{2}{*}{$>$600}  \\
                                   & $\mu\mu jj$ & 0 & 2 &    &  \\ \hline
    \multirow{2}{*}{VR} & $eejj$ & 2 & 0 & \multirow{2}{*}{$>$130}   &  \multirow{2}{*}{$<$600}  \\
                                   & $\mu\mu jj$ & 0 & 2 &    &  \\
    \hline\hline
  \end{tabular}
  \label{tab:regionDefs}
\end{table}

For the statistical analysis, a profile-likelihood fit of signal plus background templates to the data was performed using the HistFitter package~\cite{Baak:2014wma}. Systematic uncertainties, which are discussed in Section~\ref{sec:Systematics}, were incorporated into the likelihood as constrained nuisance parameters. The fit was performed in the CRs and the SR simultaneously and was used to extract  normalisation factors, i.e. scaling corrections to the event yields predicted by theoretical cross-section calculations, described in Section~\ref{sec:DataMC}.  In the CRs, only the event yield was used to extract the dominant background ($t\bar{t}$ and DY+jets) normalisation factors. In the SR, both the normalisation and the shape of $\mlqmin$ distribution were used in the fit to extract the signal normalisation factor. The templates of the $\mlqmin$ shape consisted of ten bins: six bins of 100~GeV width from 0 to 600~GeV and  four bins of 200~GeV width that cover the range up to 1.4~TeV. The template for the signal was derived from MC predictions. For the background templates, MC predictions as well as data-driven techniques were used, as detailed in the following section.

\section{Background estimation}
\label{sec:bgkEst}
Normalisation factors for the MC predictions of the two main SM backgrounds (DY+jets and \ttbar{}) were estimated with a fit to the data, as described in Section~\ref{sec:Strategy}. In total, four normalisation factors were calculated: one each for $t\bar{t}$ and  DY+jets events in both the electron and muon channels.

Smaller background contributions arise from the production of a single top quark in the $Wt$ channel, diboson events and $Z\rightarrow\tau\tau$+jets events.
These were estimated purely from simulation, i.e. their normalisation was not a free parameter in the combined fit.

Misidentified or non-prompt leptons originating from hadron decays or
photon conversions can arise in multi-jet events, single top production in
the s- or t-channel, $W$+jets and $t\bar{t}$ events (with at least one top
quark decaying hadronically). This fake-lepton background is negligible in the \mumujj~channel. In the \eejj~ channel it was evaluated using the same data-driven method as in Ref.~\cite{EXOT-2012-23}. This method was used to evaluate the migration of events among four different data samples: the nominal SR and three analogous samples selected with modified electron selection criteria.
The migration between different regions can be described by a matrix, the elements of which are functions of the proportions of true and fake electrons.
As a simplification of Ref.~\cite{EXOT-2012-23}, the fake and real rates were evaluated as a function of $\pt$~only, since they were observed to be independent of $\eta$ within the required accuracy.
They were considered to be the same for all electron candidates in an event. The fake background estimation suffers from low statistical precision.
 Its statistical uncertainty was treated as one source of the systematic uncertainty in the total background modelling.

\begin{figure}[h!]
  \centering
  \subfloat[]{
    \includegraphics[width=0.5\textwidth]{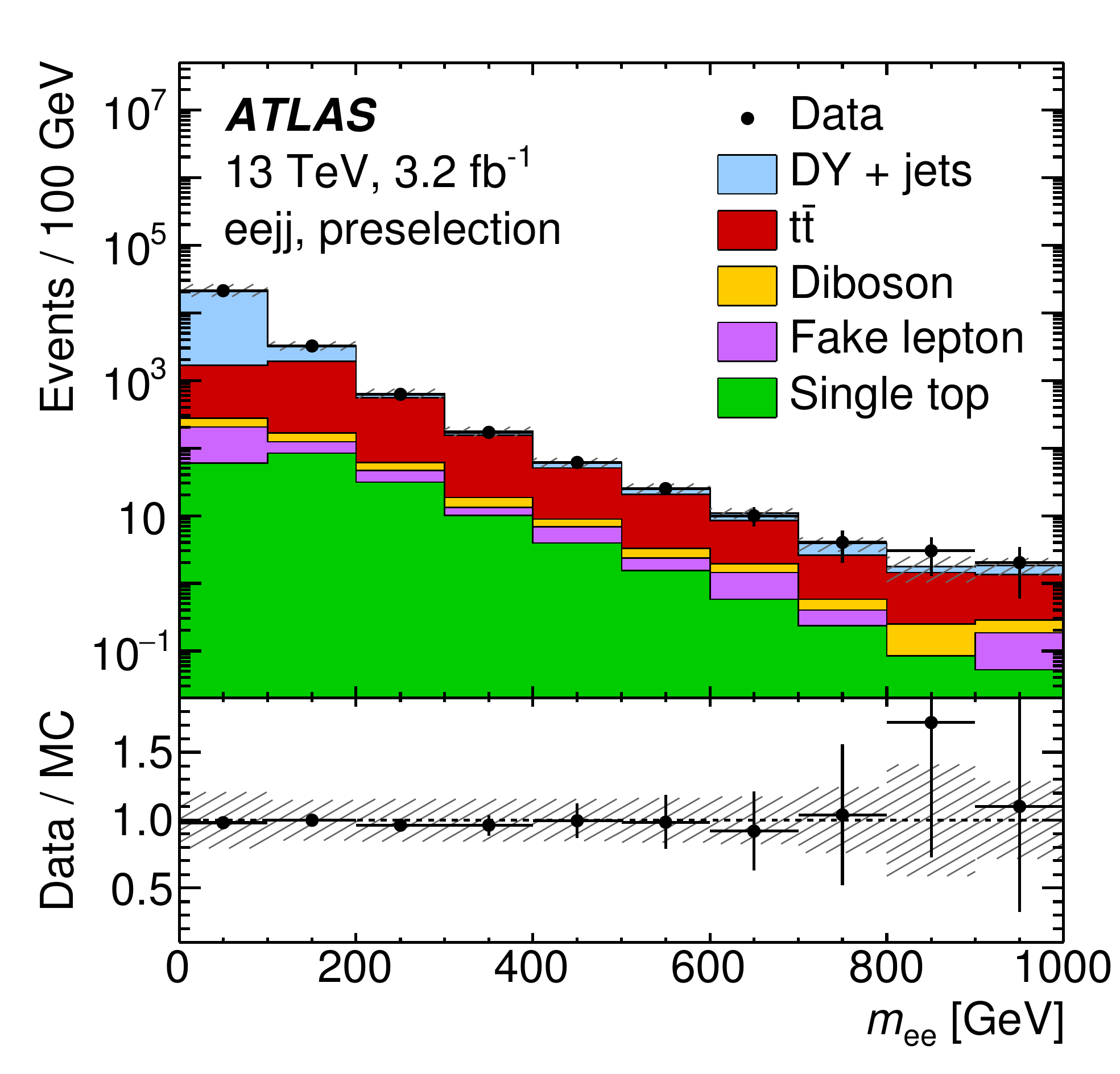}
    \label{fig:eeinvmass}
  }
  \subfloat[]{
    \includegraphics[width=0.5\textwidth]{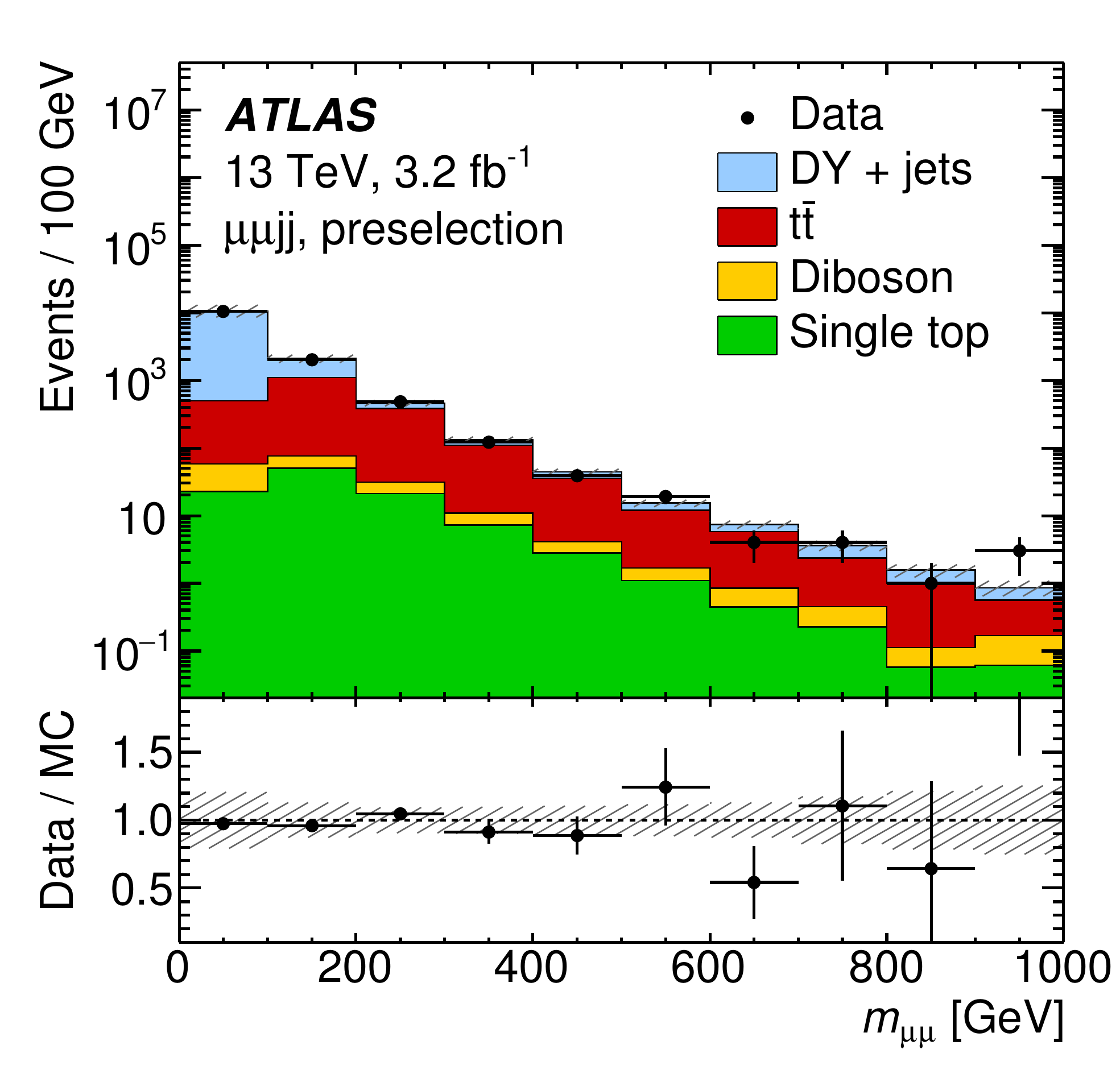}
    \label{fig:mumuinvmass}
  }
  \caption{Dilepton invariant mass for pairs of (a) electrons and (b) muons in events containing exactly two reconstructed same-flavour leptons and at least two reconstructed jets. Data are compared to the background prediction. The DY+jets and \ttbar{} expectations are shown without the normalisation factors from the fit described in Section~\ref{sec:Strategy}. The hatched bands show the total systematic uncertainty in the background prediction.}
  \label{fig:mll_presel}
\end{figure}

Figure~\ref{fig:mll_presel} shows the dilepton invariant mass for pairs of electrons (a) and muons (b) in events containing exactly two reconstructed same-flavour leptons and at least two reconstructed jets, following the selections given in Section~\ref{sec:EvtSel}. This selection stage is also referred to as preselection. The predictions of various background sources  are compared with the data. The hatched bands show the total systematic uncertainty in the background prediction.  Within the uncertainties, agreement
between data and simulation is observed. Normalisation factors for the MC
predictions for DY+jets and $t\bar{t}$ events are not applied in
the plot.

Figure~\ref{fig:ttbarcontrolregion} shows the spectrum of the minimum reconstructed lepton--jet mass in the $t\bar{t}$ control region before the fit.  Within uncertainties, the data and MC distributions are consistent.
\begin{figure}[h!]
  \centering
    \includegraphics[width=0.5\textwidth]{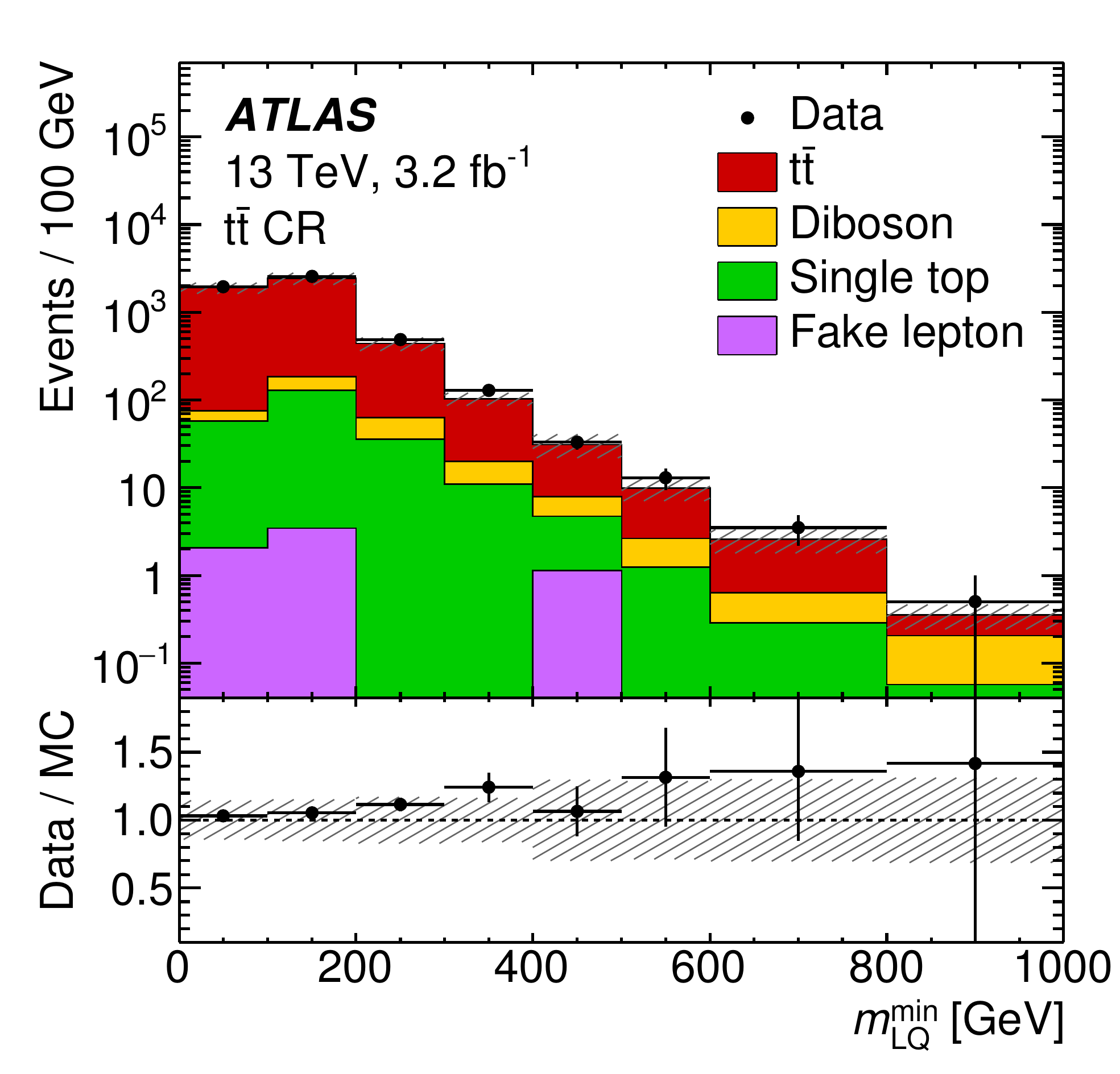}
  \caption{Minimum reconstructed lepton--jet invariant mass in the $t\bar{t}$ control region. Data are compared to the background prediction. The \ttbar{} prediction is shown without the normalisation factor from the fit described in Section~\ref{sec:Strategy}. The hatched bands show the total systematic uncertainty
in the background prediction.}
  \label{fig:ttbarcontrolregion}
\end{figure}

\section{Sources of systematic uncertainties}\label{sec:syst}
\label{sec:Systematics}
The following sources of systematic uncertainty were considered:
\begin{itemize}
\item The uncertainty in the integrated luminosity is $5$\%. It was derived following a methodology similar to that detailed in Ref.~\cite{DAPR-2011-01},
from a calibration of the luminosity scale using $x$--$y$ beam-separation scans performed in August 2015. This uncertainty affects the predicted signal event yield and those background rates for which theoretical estimates are used.
\item The jet energy scale (JES) uncertainty depends on the \pt~and \eta of the jet and on the pile-up conditions in an event.
 A further uncertainty in the jet energy resolution (JER) was taken into account. These sources each correspond to uncertainties in the jet energy of up to 3\%. 
The largest resulting uncertainty in the background event yields is about 10\% in the control regions. In the signal region, the uncertainty in the event yields amounts to at most 5\% for the background and less than 1\% for the signal.
\item The uncertainty in the lepton trigger efficiency scale factors is around $2\%$ for the kinematic region considered here. 
\item Differences between the MC and data in the efficiency of the isolation requirement on the selected muons correspond to uncertainties of
1--5\% on the scale factor to correct the MC prediction. Other muon-related
uncertainties arise from the momentum scale, resolution, and quality criteria and typically affect the muon event yields by around $1\%$ in all regions for both signal and background.
\item Uncertainties in the electron energy scale, identification and isolation affect the electron event yields by up to 2\% in all regions for both signal and background.
\item Uncertainties due to choices that have to be made in the event generation which affect final-state observables were estimated for the two major background sources: $t\bar{t}$ and DY+jets. These \emph{modelling uncertainties} refer to e.g. possible differences in the generation of the hard scattering, scale dependencies,  the parton shower and hadronisation and fragmentation models.
Differences in the background modelling can change the event yields (total normalisation) in the CRs. 
Moreover, there is an uncertainty in the shape of the \mlqmin{} distribution in the signal region due to background modelling effects. This was
estimated as one uncertainty per \mlqmin{}-bin in the signal region and propagated to the normalisation factor by the fit. The uncertainties were treated as uncorrelated between different bins.\\
The impact of modelling uncertainties in final-state predictions for $t\bar{t}$ processes was quantified by comparing various simulated samples: differences due to the parton shower as well as the hadronisation and fragmentation model were estimated by comparing the nominal sample to one that uses \textsc{Herwig++}~\cite{Corcella:2000bw}, effects of additional or reduced radiation were estimated by varying the parton-shower and scale parameters within \pythia, and an alternative generator (\textsc{aMC@NLO}~\cite{Alwall:2014hca} with \textsc{Herwig++}) was used to estimate differences in the hard scatter generation.
 The total uncertainty in the predicted \ttbar{} event yield varies between $14\%$ (in the $t\bar{t}$ CR) to about $30\%$ (in the signal regions).\\ 
 Modelling uncertainties for the DY+jets background were assessed with different approaches, simulation-based, as well as data-driven in different regions of phase space.
The baseline estimate used events from the DY+jets control region with \st{} higher than 600~GeV. In this region, the shapes of both the \mlqmin{} and \st{} distributions are very similar to those in the signal region, which differs only by the cut on the dilepton invariant mass. This cut was found to not affect the shapes of the other discriminating variables. The difference between the data and the background prediction in this region was used as an estimate of the modelling uncertainty. \\
The result was cross-checked using simulated samples in which the renormalisation, factorisation and  resummation scales, as well as the scale for matrix element and parton shower matching, were independently varied up and down by a factor of two. Within the statistical uncertainties resulting from the limited number of events in these samples, the estimate from the data-driven approach was confirmed. The result is a 10\% uncertainty in the DY+jets event yield in the control regions and 20\% in each \mlqmin{} bin  in the signal region.
\item PDF uncertainties on the \ttbar{} and DY+jets normalisation as well as the \mlqmin{} shape amount to less than 4\% and do not affect the final result given the large modelling uncertainties described above.
\item The effects of higher-order contributions on the signal cross section were estimated by varying the QCD renormalisation and factorisation scales, set to a common value, up and
down by a factor of two, as done in Ref.~\cite{Kramer:2004df}.  One half of the difference between the predicted cross section for the increased and reduced scale choice is used as
the cross section uncertainty for a given mass. This uncertainty lies in the  the range 12--17\% for the mass points considered in this paper.
\item The impact of theoretical uncertainties related to the parton-shower algorithm and multiple-interactions tune were evaluated by varying the
corresponding parameters, as specified in Ref.~\cite{ATL-PHYS-PUB-2011-014}. This leads to an uncertainty in signal acceptance of up to 2\%.
\item
The uncertainty in the signal cross section due to the choice of PDF set was calculated as the envelope of the predictions of 40 different \textsc{CTEQ6.6}~NLO error sets~\cite{Kramer:2004df}. The uncertainty ranges from 11\% at $m_{\mathrm{LQ}}$\ = 500 \GeV\ to 34\% at $m_{\mathrm{LQ}}$\ = 1500 \GeV.
The predicted signal acceptance was studied using
 \textsc{NNPDF23LO}~\cite{Ball:2012cx}, \textsc{CT14}~\cite{Dulat:2015mca} and \textsc{MMHT14}~\cite{Harland-Lang:2014zoa} PDF sets. The acceptance is very insensitive to the choice
of parton distribution function; the systematic uncertainty from this source is less than 1\%.
\end{itemize}

\section{Results}\label{sec:res}
The results are consistent with SM expectations. The normalisation factors obtained in the fit described in Section~\ref{sec:Strategy} are summarised in Table~\ref{tab:normFacts}. Similar results are obtained in the two channels and the normalisation factors are found to be compatible with unity within the uncertainties.
\begin{table}[tb]
\caption{Normalisation factors for the main backgrounds obtained from the combined fit in each of the channels.}
\begin{center}
\begin{tabular}{c|c|c}
\midrule \midrule
Channel     & DY+jets     & \ttbar{}    \\
\toprule
$eejj$      & $0.9\pm 0.1 $ & $1.0 \pm 0.1$ \\
$\mu\mu jj$ & $0.9 \pm 0.1 $ & $1.0^{+0.2}_{-0.1}$ \\
\midrule \midrule
\end{tabular}
\end{center}
\label{tab:normFacts}
\end{table}
The reliability of extrapolating the background predictions from the control regions to the signal region was checked in the VRs defined in Section~\ref{sec:Strategy}.
The distribution of $\mlqmin$ in the two validation regions is compared to
the background prediction after the fit  in Figure~\ref{fig:vrmlqmin}.
Within the uncertainties, the predictions are compatible with the observed data.
\begin{figure}[htbp]
  \centering
  \subfloat[]{
    \includegraphics[width=0.5\textwidth]{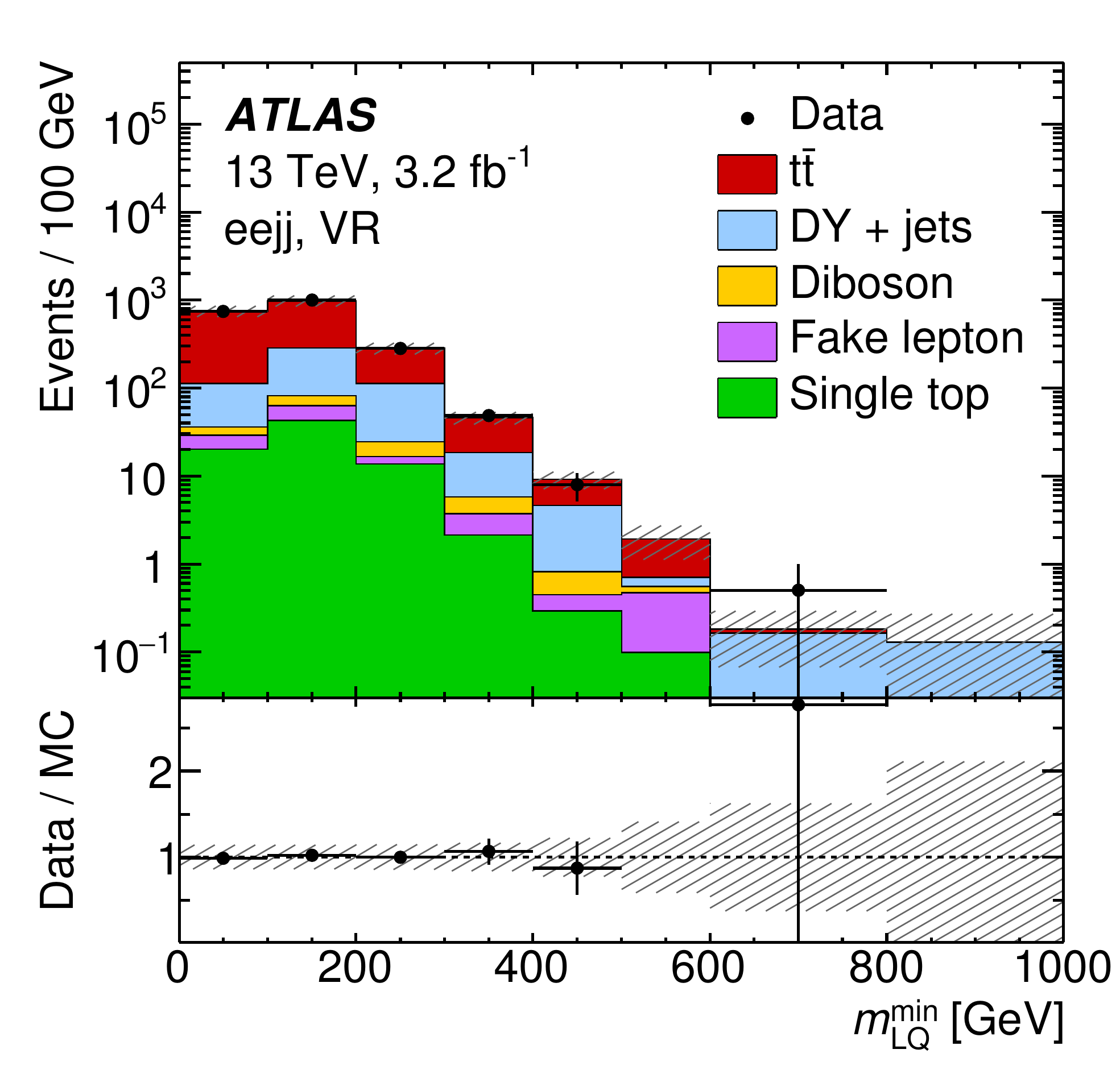}
    \label{fig:eeinvmass}
  }
  \subfloat[]{
    \includegraphics[width=0.5\textwidth]{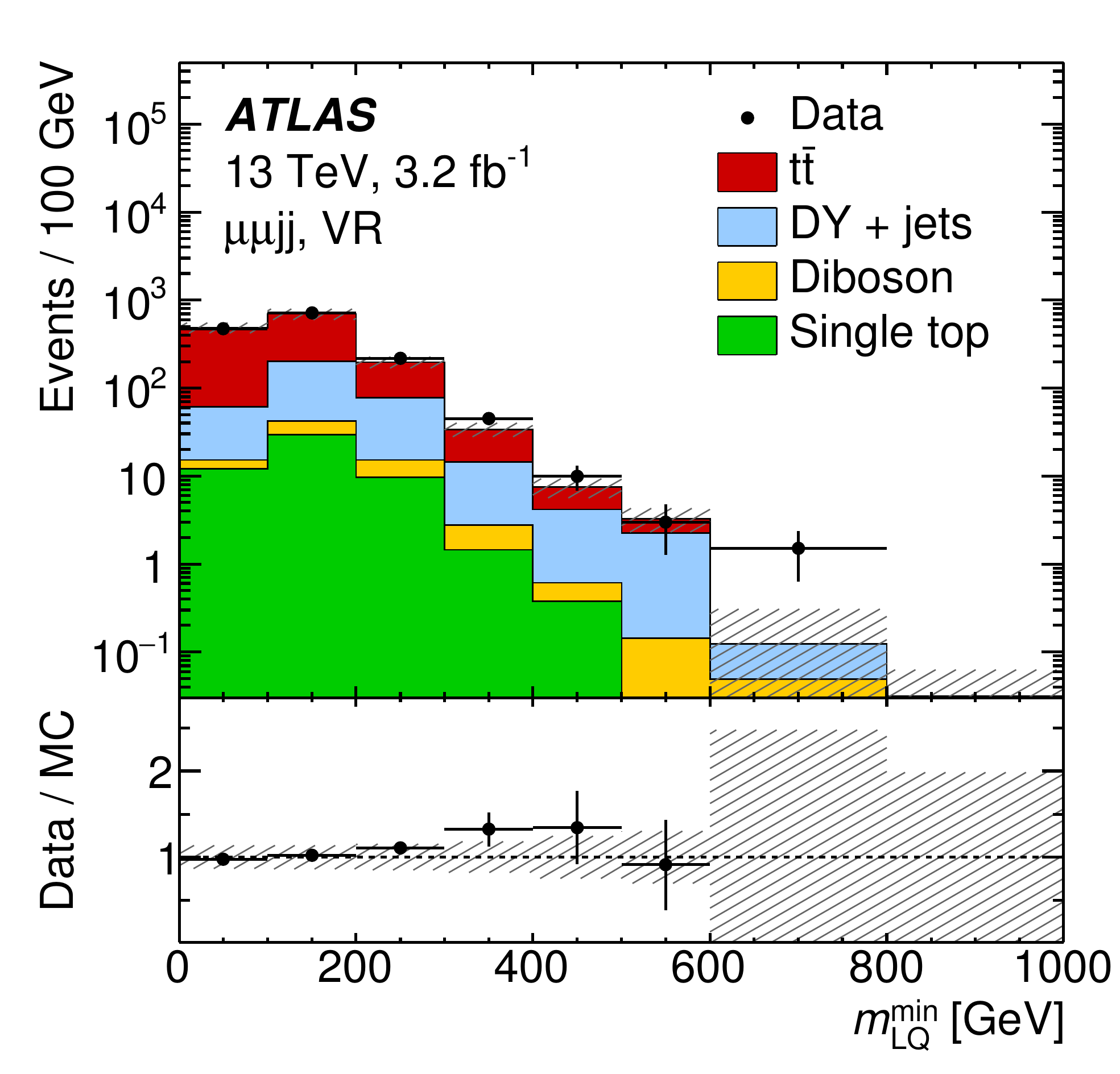}
    \label{fig:mumuinvmass}
  }
  \caption{Minimum reconstructed lepton--jet invariant mass in the validation regions for pairs of (a) electrons and (b) muons. Data are compared to the background prediction. The DY+jets and \ttbar{} expectations are scaled by the normalisation factors obtained from the fit described in Section~\ref{sec:Strategy}. The hatched bands show the total systematic uncertainty
in the background prediction.  }
  \label{fig:vrmlqmin}
\end{figure}

The observed and expected event yields in the signal regions for the \eejj and
the \mumujj channels, after the fits, are shown in Tables~\ref{table_eejj}
and \ref{table_mumujj}, respectively.  The values in these tables are
intended to illustrate the sensitivity independent of a specific signal
hypothesis and thus, in this case, the fit was performed using only the
control regions as input.
The resulting fit parameters (DY+jets and \ttbar{} normalisation factors and values of the nuisance parameters) were transferred to the signal region, using appropriate transfer factors based on the MC models. The different contributions to the background do not necessarily exactly sum to the total quoted number of background events owing to the rounding scheme used.
The dominant experimental systematic uncertainties in the background prediction arise from uncertainties on corrections to the simulated electron and muon trigger efficiencies and the jet energy scale; the latter source gives an uncertainty of 2--4\% in the signal region.
The luminosity uncertainty of 5\% contributes for simulated backgrounds not constrained by the fit (diboson and single-top production).
The theoretical uncertainties after the fit range from 3\% to 12\% for the DY+jets background and from 3\% to 16\% for the \ttbar{} background. The global fit takes correlations of the nuisance parameters into account, which results in the uncertainty in the total background being smaller than the quadratic sum of the uncertainties in the separate components. The contribution of $Z\rightarrow \tau \tau +$jets to the background is negligible and not shown in the tables.
\begin{table}[tb]
\caption{Observed and predicted event yields in the signal and control regions in the \eejj channel. The background prediction with its total uncertainty after the fit is shown. The fit is performed using only the control regions as input. The lower part of the table shows the separate contributions from the different background processes and their total uncertainty after the fit. In addition, the expected signal event yields for $\beta=1$ and $\LQ$ masses of 500,1000, and 1500~GeV are given.}
\begin{center}
\setlength{\tabcolsep}{0.0pc}
{\small
\begin{tabular*}{\textwidth}{@{\extracolsep{\fill}}lrrrr}
\noalign{\smallskip}\hline\hline\noalign{\smallskip}
                                                & SR           & CR DY+jets      & CR \ttbar{}   \\[-0.05cm]
\noalign{\smallskip}\toprule\noalign{\smallskip}
Observed events                                 & $279$        & $20328$          & $5194$         \\
\noalign{\smallskip}\midrule\noalign{\smallskip}
Total background events                        & $300 \pm 30$ & $20300 \pm 200$  & $5200 \pm 50$  \\
\noalign{\smallskip}\midrule\noalign{\smallskip}
        Fitted DY+jets events                   & $74 \pm 7$  & $19100 \pm 200$ & $<0.01$  \\
        Fitted \ttbar{} events                  & $190 \pm 30$ & $1060 \pm 10$    & $4840 \pm 40$ \\
        MC predicted diboson events                   & $12.5 \pm 0.6$  & $63  \pm 3$     & $115 \pm 6$    \\
        MC predicted single-top events                & $20 \pm 1$   & $42 \pm 2$       & $230 \pm 10$   \\
        Estimated fake-lepton events               & $9 \pm 4$   & $120 \pm 10$    & $6 \pm 3$   \\ \midrule
       MC exp. signal events ($\mlq=500$~GeV)  & $1000 \pm 100$     & $26 \pm 4$          & $<0.01$        \\
       MC exp. signal events ($\mlq=1000$~GeV) & $13 \pm 2$     & $0.03 \pm 0.00$          & $<0.01$        \\
       MC exp. signal events ($\mlq=1500$~GeV) & $0.6 \pm 0.1$     & $<0.01$          & $<0.01$        \\

\noalign{\smallskip}\midrule\midrule\noalign{\smallskip}
\end{tabular*}
}
\end{center}
\label{table_eejj}
\end{table}
\begin{table}[tb]
\caption{Observed and predicted event yields in the signal and control regions in the \mumujj channel. The background prediction with its total uncertainty after the fit is shown. The fit is performed using only the control regions as input. The lower part of the table shows the separate contributions from the different background processes and their total uncertainty after the fit. Where no number is given, the contribution is found to be negligible. In addition, the expected signal event yields for $\beta=1$ and $\LQ$ masses of 500,1000, and 1500~GeV are given.}
\begin{center}
\setlength{\tabcolsep}{0.0pc}
{\small
\begin{tabular*}{\textwidth}{@{\extracolsep{\fill}}lrrrrrrrrrrrr}
\noalign{\smallskip}\hline\hline\noalign{\smallskip}
                                               & SR           & CR DY+jets      & CR \ttbar{}                \\[-0.05cm]
\noalign{\smallskip}\toprule\noalign{\smallskip}
Observed events                                & $188$        & $10233$         & $5194$                     \\
\noalign{\smallskip}\midrule\noalign{\smallskip}
Fitted background events                       & $200 \pm 30$ & $10200 \pm 100$ & $5200 \pm 70$              \\
\noalign{\smallskip}\midrule\noalign{\smallskip}
        Fitted DY+jets events                  & $56 \pm 8$   & $9800 \pm 100 $ & $9 \pm 1 $                 \\
        Fitted \ttbar{} events                 & $120 \pm 30$ & $400 \pm 20$    & $4840 \pm 80 $ &           \\
        MC predicted diboson events                  & $8.6 \pm 0.6$   & $32 \pm 3$      & $115 \pm 10$               \\
        MC predicted single-top events                & $12.8 \pm 0.9$   & $18 \pm 2$     & $230 \pm 20$               \\
         Estimated fake-lepton events              & --           & --              & --                         \\ \midrule
       MC exp. signal events ($\mlq=500$~GeV) & $610 \pm 40$     & $25 \pm 2$            & $3 \pm 3$              \\
       MC exp. signal events ($\mlq=1000$~GeV) & $8.0 \pm 0.8$     & $0.08 \pm 0.01$         & $0.1 \pm 0.1$            \\
       MC exp. signal events ($\mlq=1500$~GeV) & $0.33 \pm 0.06$     & $<0.01$         & $<0.01$
\\
\noalign{\smallskip}\midrule\midrule\noalign{\smallskip}
\end{tabular*}
}
\end{center}
\label{table_mumujj}
\end{table}

Figure~\ref{fig:SR} shows the SR distribution of $\mlqmin$ compared to background
 predictions based on the combined fit in the CRs and the SR. The signal prediction for a LQ of mass 1.1~TeV is also shown. 
The wider signal shape in the muon channel compared to the electron channel is due the worsening of the muon momentum resolution with increasing momentum.
\begin{figure}[htbp]
  \centering
  \subfloat[]{
    \includegraphics[width=0.5\textwidth]{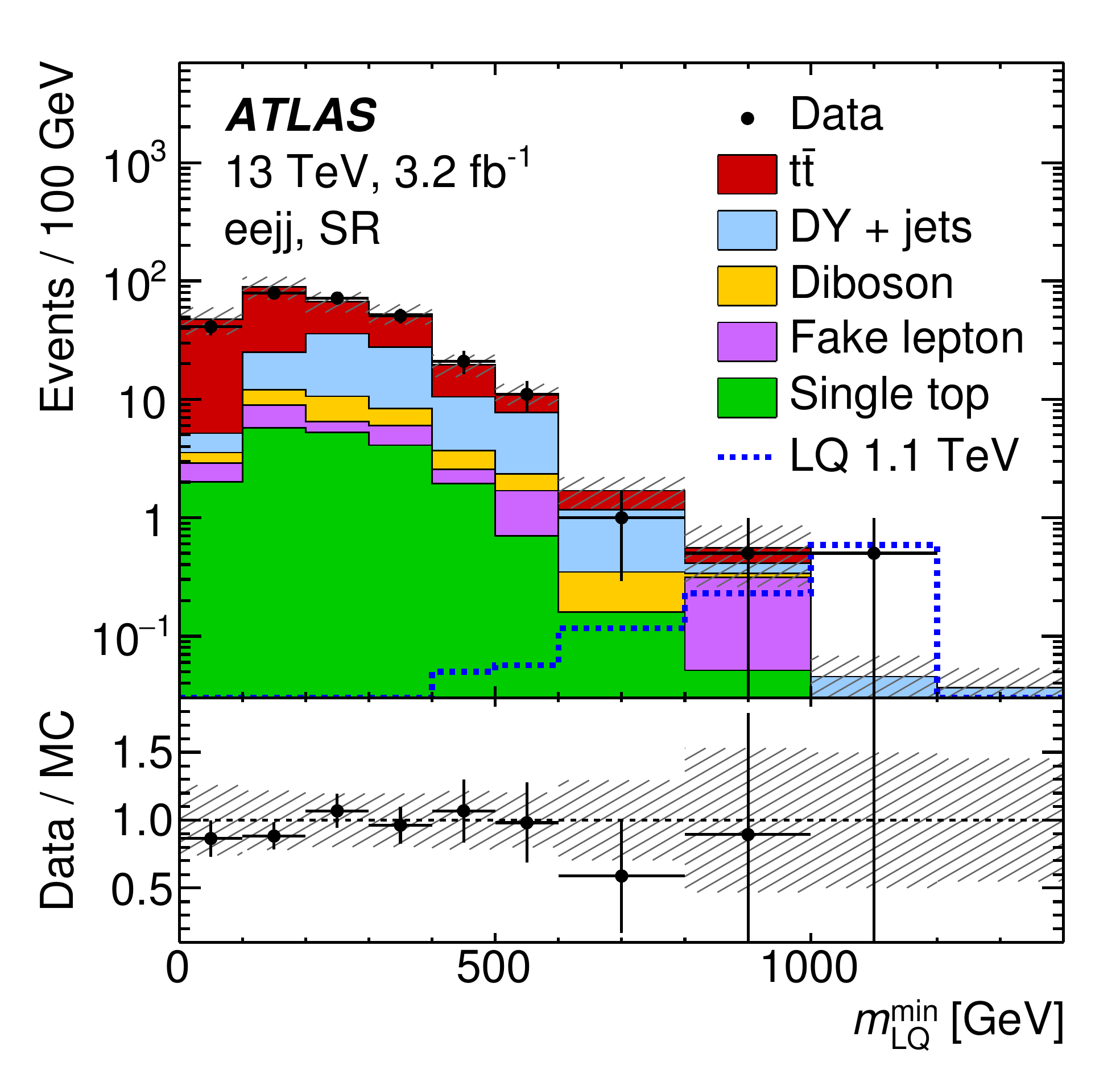}
  }
  \subfloat[]{
    \includegraphics[width=0.5\textwidth]{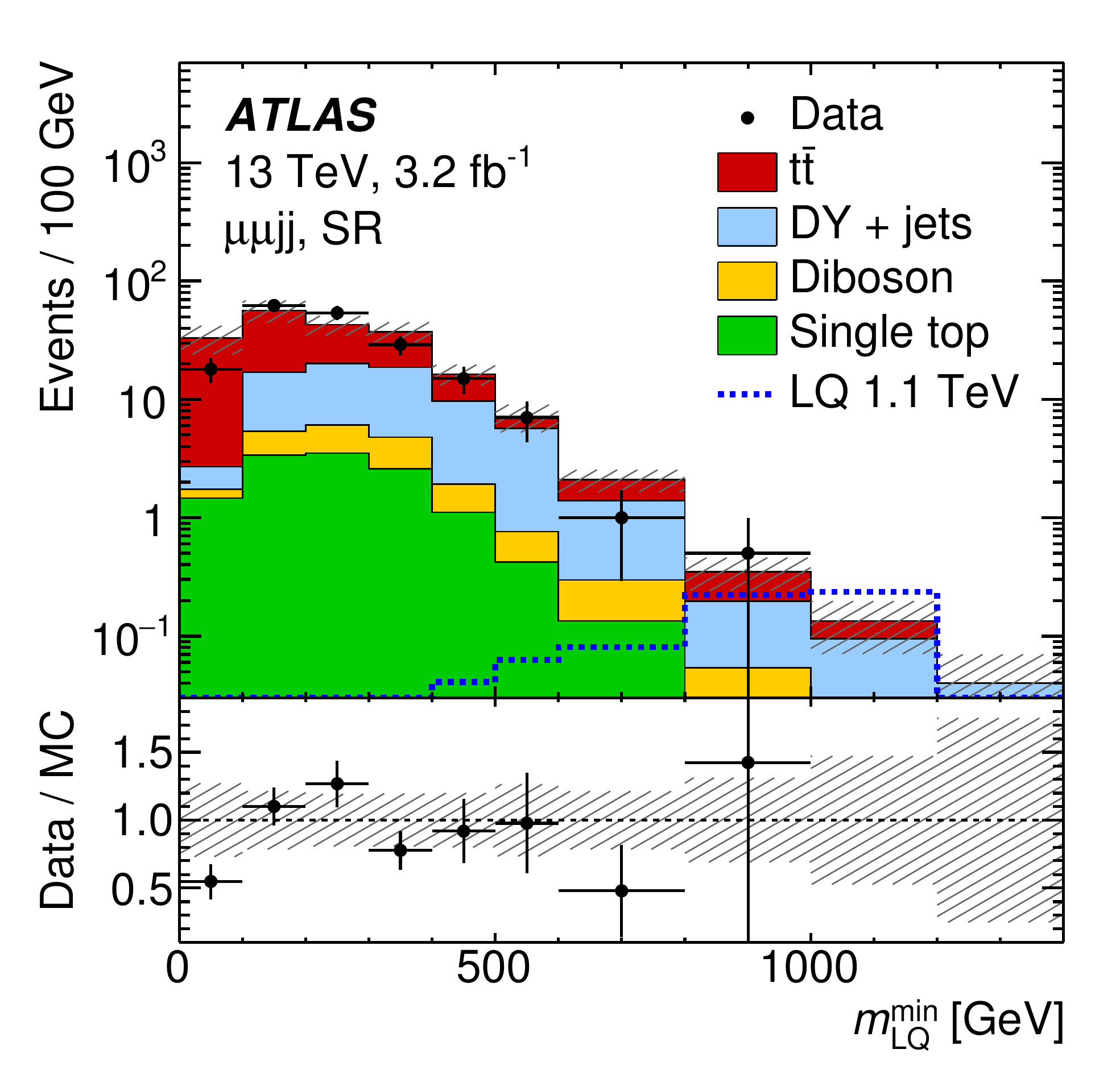}
  }
  \caption{Distribution of the minimum reconstructed $\LQ$ candidate mass, $\mlqmin$, in the signal region of (a) the first-generation leptoquark search and (b) the second-generation search. Data are compared to the background prediction. The DY+jets and \ttbar{} expectations are scaled by the normalisation factors obtained from the fit described in Section~\ref{sec:Strategy}. The signal expectation for a $\LQ$ of mass 1.1~TeV is also shown. The hatched bands show the total systematic uncertainty
in the background prediction. }
  \label{fig:SR}
\end{figure}
Again, no significant deviation from the SM predictions was observed.
Limits on the LQ signal strength were derived using pseudo-experiments and following a modified frequentist ${\mathrm{CL}}_s$ method \cite{0954-3899-28-10-313}.
Apart from the luminosity uncertainty (5\%), dominant uncertainties in the signal event yields arise from lepton scale factors and are of the order of 2--5\% at low masses and up to 10\% at high masses.

In Figure~\ref{fig:limits}, limits on the cross section times branching ratio are shown on the left, for leptoquarks of the first (second) generation in the top (bottom) plot.
The expected limit is depicted by the dashed line; the uncertainty bands result from considering all sources of systematic as well as statistical uncertainties.
The observed limit is given by the solid line.
The NLO pair-production cross section for $\beta=1$ is shown as a line with a shaded band representing the uncertainties. The shaded band around it illustrates the uncertainties in the theoretical prediction due to PDF and scale uncertainties.
The intersection of this line with the cross-section limits yields the lower limit on the leptoquark mass for a value of $\beta=1$.
These observed (expected) limits are found to be 1100~GeV (1160~GeV) and 1050~GeV (1040~GeV)  for first- and second-generation LQs, respectively. The observed limit for each channel is
stronger than the previous bound~\cite{EXOT-2014-03} by 50~GeV. The expected limits are improved by 110~GeV for the first-generation search and by
40~GeV in the second generation search.
The theoretical cross section was scaled by $\beta^2$ and then used to obtain the limits on the
branching ratio as a function of the LQ mass shown on the right of Figure~\ref{fig:limits}. Below $\LQ${} masses of 650~GeV, the limits on $\beta$ are weaker than those obtained at 8~TeV centre-of-mass energy, which are shown as the dashed-dotted line, owing to the much lower integrated luminosity collected in 2015 and the effects of background at lower LQ masses. At high masses (above 900~GeV), however, the gain in the production cross section at 13~TeV compensates for the smaller luminosity and stronger bounds than at 8~TeV are obtained. In the intermediate mass region, the results are comparable. Mass limits for various values of $\beta$ are summarised in Table~\ref{tab:massLimits}.

\begin{figure}[h!]
  \centering
  \subfloat[]{
    \includegraphics[width=0.5\textwidth]{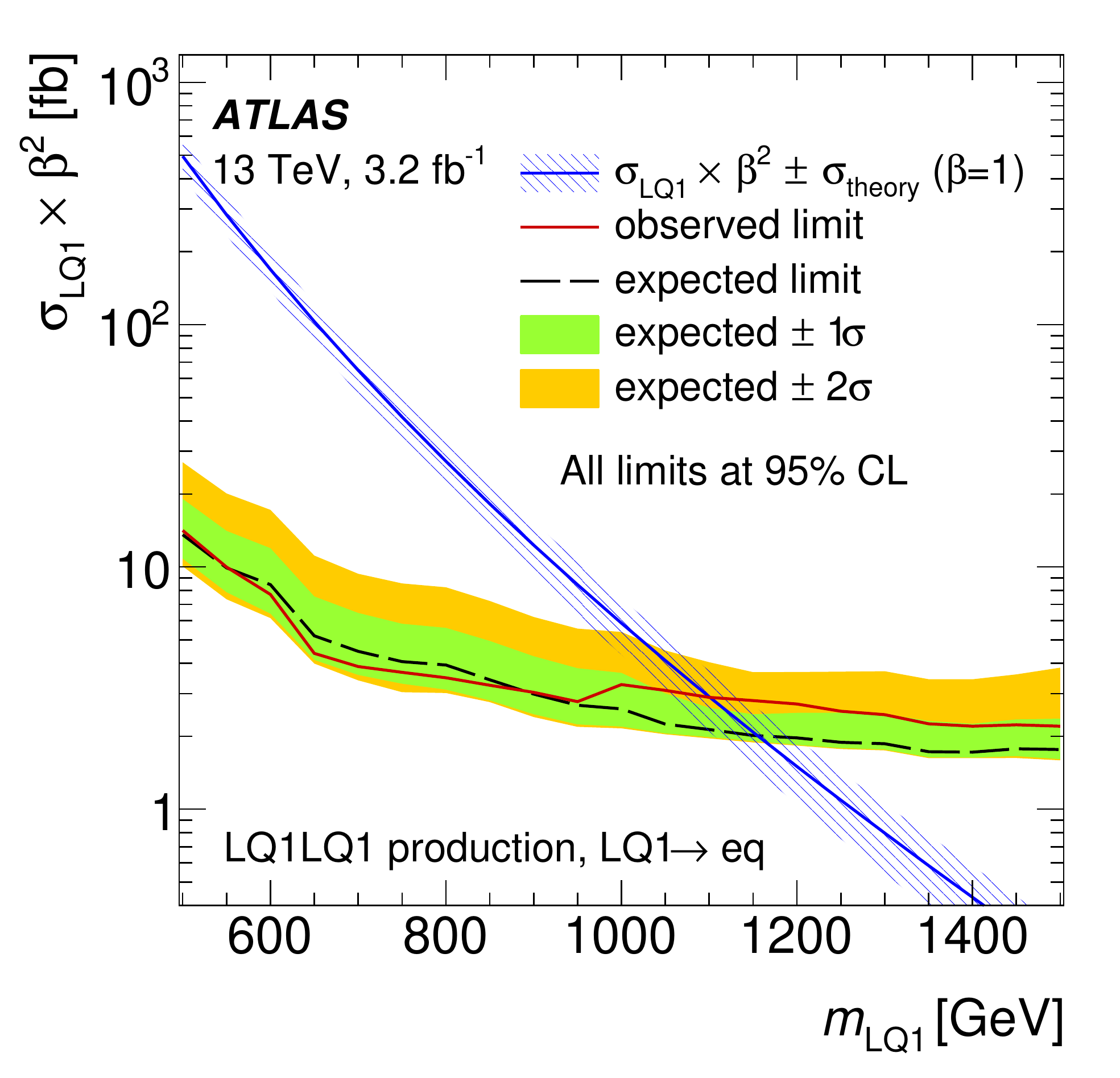}
    \label{fig:eejjlim}
  }
  \subfloat[]{
    \includegraphics[width=0.5\textwidth]{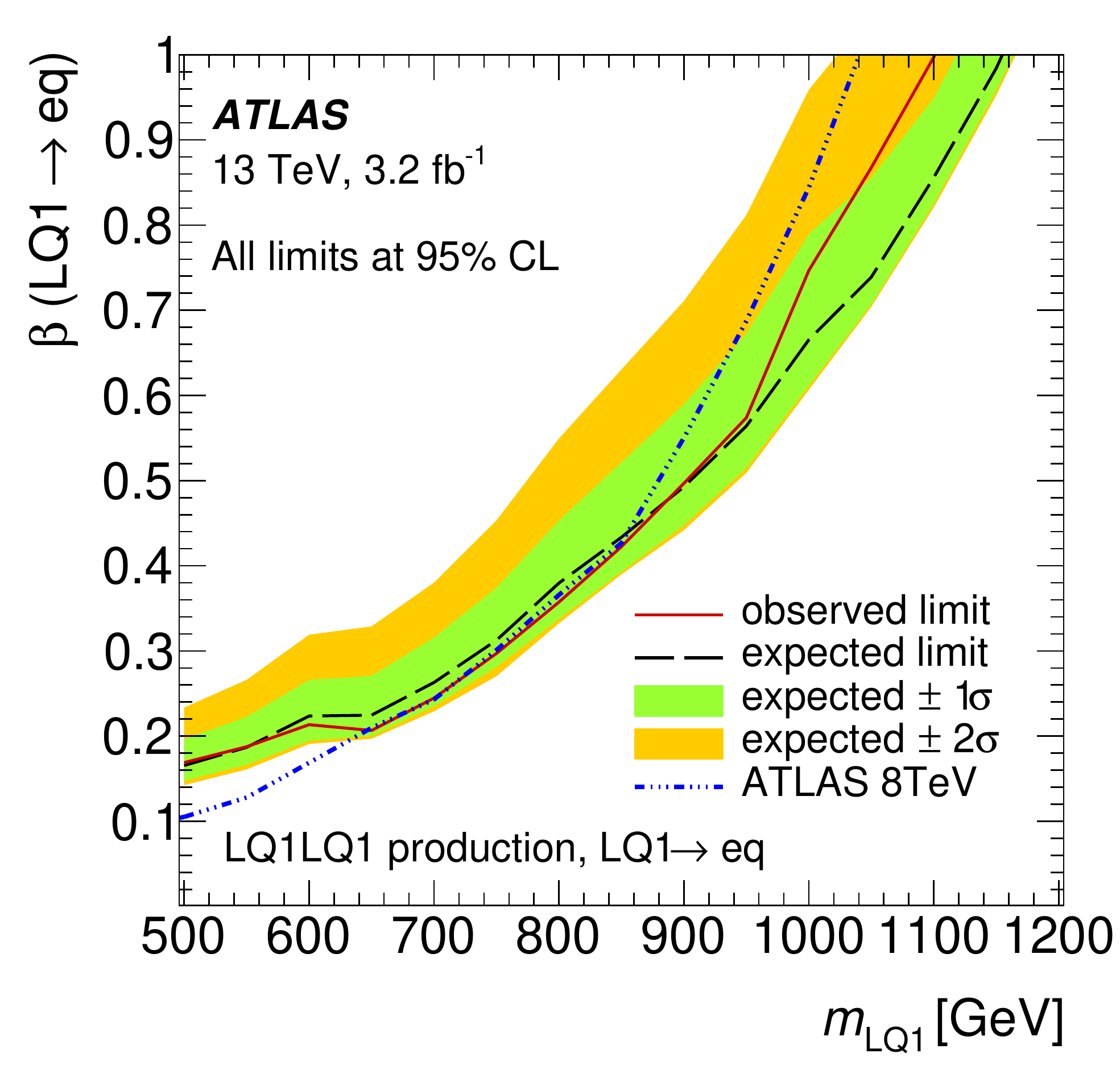}
    \label{fig:eejjbeta}
  }
  \\
    \subfloat[]{
    \includegraphics[width=0.5\textwidth]{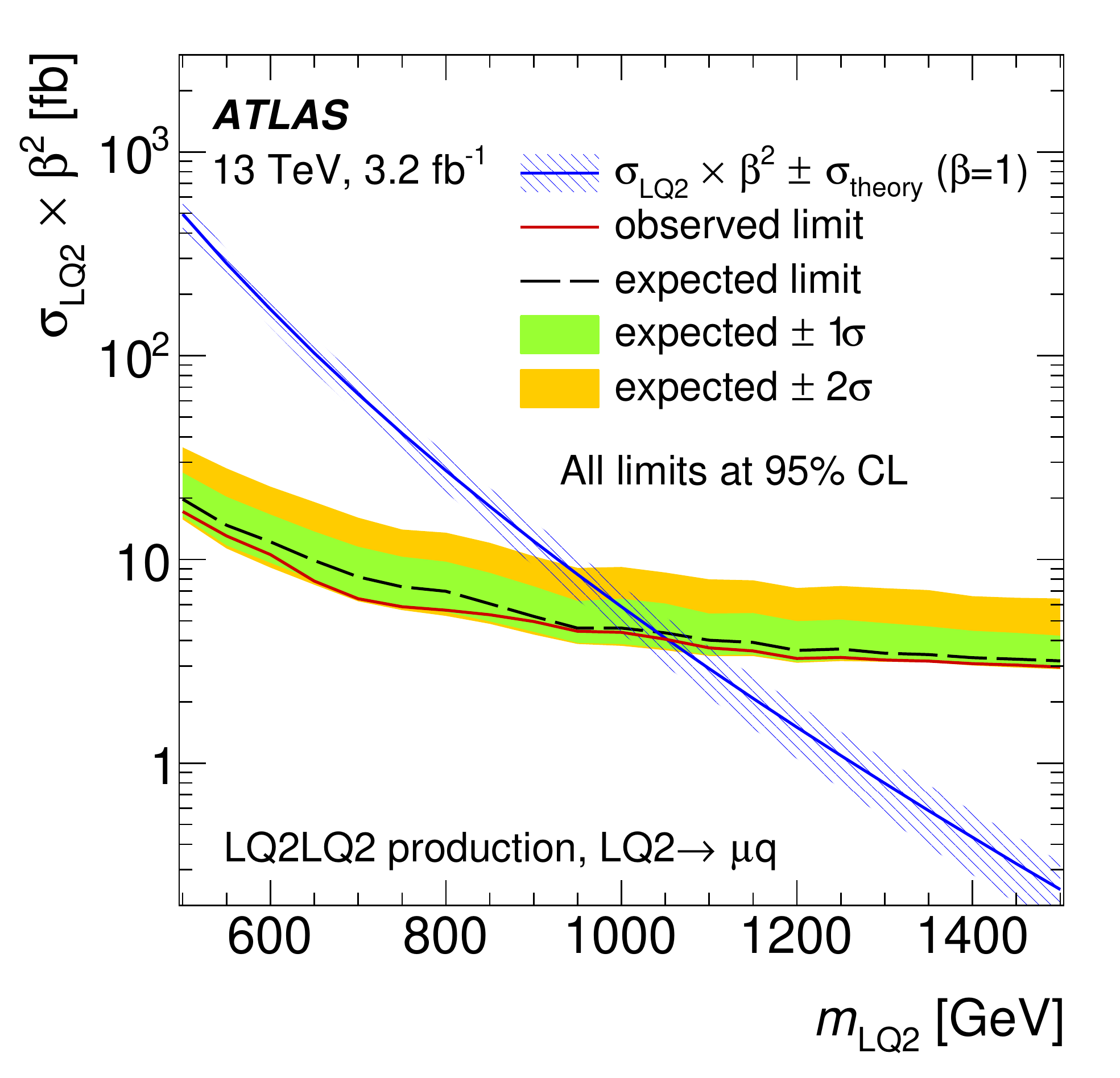}
    \label{fig:mumujjlim}
  }
  \subfloat[]{
    \includegraphics[width=0.5\textwidth]{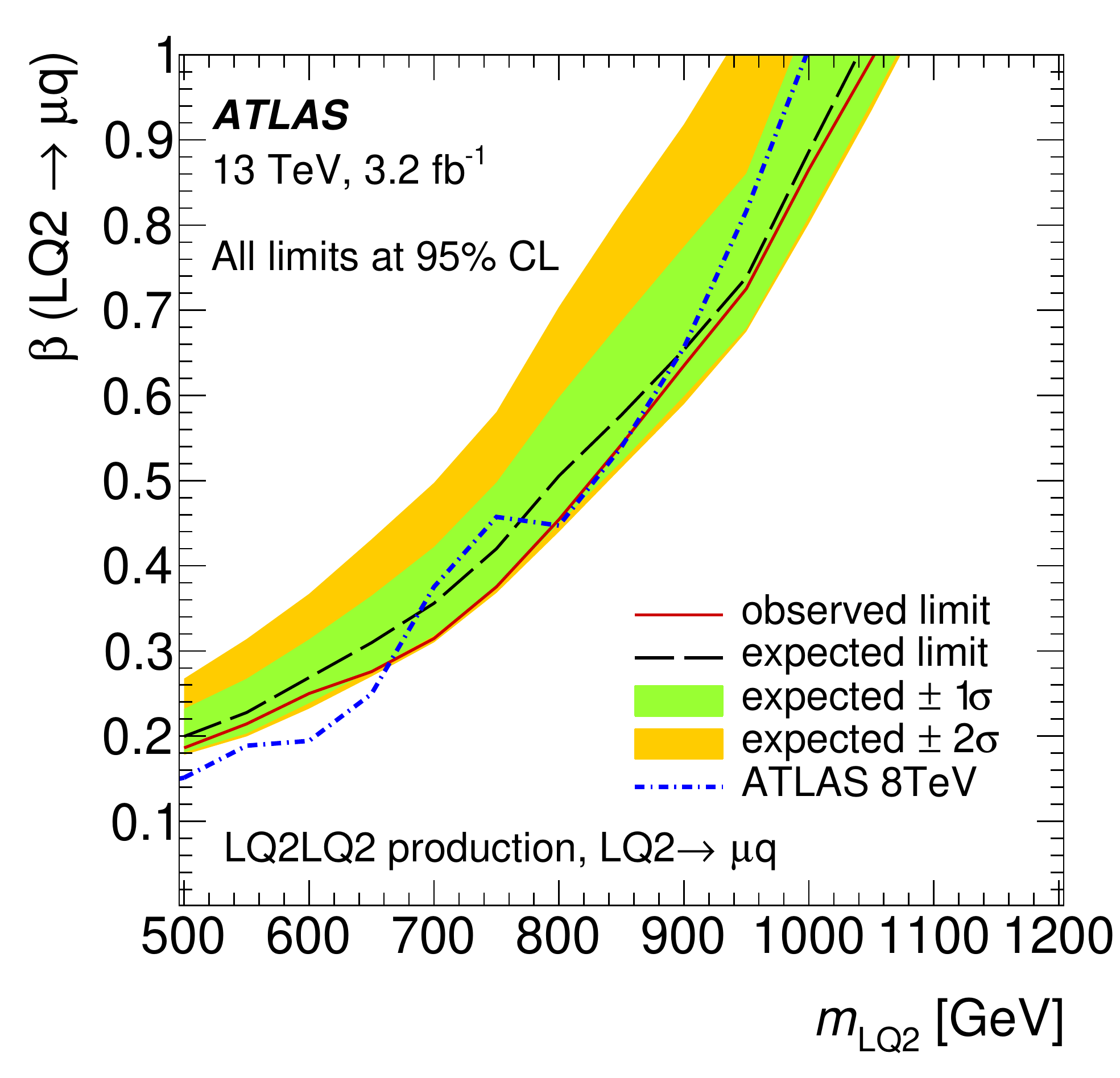}
    \label{fig:mumujjbeta}
  }
  \caption{The cross-section limits (a) on scalar LQ pair production times the square of the branching ratio as a function of mass and (b) limits on the branching ratio as a function of mass for first-generation leptoquarks. Analogous limits are shown for second-generation leptoquarks in (c) and (d). Expected and observed limits are also shown. The uncertainty bands on the expected limit represent all sources of systematic and statistical uncertainty.  On the left, the expected NLO production cross section ($\beta$\ = 1.0) for scalar leptoquark pair-production and its corresponding theoretical uncertainty due to the choice of PDF set and renormalisation/factorisation scale are also included.  The observed limit from the 8~TeV analysis is also shown on plots (b) and (d)~\cite{EXOT-2014-03}. }
  \label{fig:limits}
\end{figure}
\FloatBarrier

\begin{table}[htb]
\caption{Expected and observed 95\%\,CL lower limits on first- and second-generation leptoquark masses for different assumptions of $\beta$.}
\begin{center}
\begin{tabular}{c|cc|cc}
\hline \hline
\multirow{3}{*}{$\beta$} & \multicolumn{4}{c}{95\% CL limit on}                                                     \\
                         & \multicolumn{2}{c|}{$m_{\text{LQ1}}$ [GeV]} & \multicolumn{2}{c}{$m_{\text{LQ2}}$ [GeV]} \\
                         & Expected                                    & Observed & Expected & Observed             \\
\toprule
1.00                     & 1160                                        & 1100     & 1040     & 1050                 \\
0.75                     & 1050                                        & 1000      & 950      & 960                  \\
0.50                     & 900                                         & 900      & 800      & 830                  \\
0.25                     & 680                                         & 700      & 580      & 600                  \\
\hline \hline
\end{tabular}
\end{center}
\label{tab:massLimits}
\end{table}

\FloatBarrier

\section{Summary and conclusions}
Searches for first- and second-generation scalar leptoquarks, pair-produced in $pp$ collisions at 13~TeV centre-of-mass energy, have been performed with the ATLAS detector
at the LHC. An integrated luminosity of 3.2~\invfb~of data was used.  No significant excess above the SM background expectation was observed in either channel.
The results were interpreted in the framework of the mBRW model.
Mass-dependent limits were derived on the pair-production cross section times the square
 of the branching ratio ($\beta^2$) and on $\beta$. For $\beta=1$, the observed (expected) $\LQ${} mass limits at 95\%\,CL are 1100~GeV and 1050~GeV (1160~GeV and 1040~GeV) for
first- and second-generation leptoquarks, respectively. The observed bounds are more stringent than the previous ATLAS limits by 50~GeV in each channel. This analysis is
the first result at 13~TeV using the Run 2 data collected by the ATLAS experiment in a program of high precision inclusive searches for resonant signatures involving a
lepton and a jet.

\section*{Acknowledgements}

We thank CERN for the very successful operation of the LHC, as well as the
support staff from our institutions without whom ATLAS could not be
operated efficiently.

We acknowledge the support of ANPCyT, Argentina; YerPhI, Armenia; ARC, Australia; BMWFW and FWF, Austria; ANAS, Azerbaijan; SSTC, Belarus; CNPq and FAPESP, Brazil; NSERC, NRC and CFI, Canada; CERN; CONICYT, Chile; CAS, MOST and NSFC, China; COLCIENCIAS, Colombia; MSMT CR, MPO CR and VSC CR, Czech Republic; DNRF and DNSRC, Denmark; IN2P3-CNRS, CEA-DSM/IRFU, France; GNSF, Georgia; BMBF, HGF, and MPG, Germany; GSRT, Greece; RGC, Hong Kong SAR, China; ISF, I-CORE and Benoziyo Center, Israel; INFN, Italy; MEXT and JSPS, Japan; CNRST, Morocco; FOM and NWO, Netherlands; RCN, Norway; MNiSW and NCN, Poland; FCT, Portugal; MNE/IFA, Romania; MES of Russia and NRC KI, Russian Federation; JINR; MESTD, Serbia; MSSR, Slovakia; ARRS and MIZ\v{S}, Slovenia; DST/NRF, South Africa; MINECO, Spain; SRC and Wallenberg Foundation, Sweden; SERI, SNSF and Cantons of Bern and Geneva, Switzerland; MOST, Taiwan; TAEK, Turkey; STFC, United Kingdom; DOE and NSF, United States of America. In addition, individual groups and members have received support from BCKDF, the Canada Council, CANARIE, CRC, Compute Canada, FQRNT, and the Ontario Innovation Trust, Canada; EPLANET, ERC, FP7, Horizon 2020 and Marie Sk{\l}odowska-Curie Actions, European Union; Investissements d'Avenir Labex and Idex, ANR, R{\'e}gion Auvergne and Fondation Partager le Savoir, France; DFG and AvH Foundation, Germany; Herakleitos, Thales and Aristeia programmes co-financed by EU-ESF and the Greek NSRF; BSF, GIF and Minerva, Israel; BRF, Norway; Generalitat de Catalunya, Generalitat Valenciana, Spain; the Royal Society and Leverhulme Trust, United Kingdom.

The crucial computing support from all WLCG partners is acknowledged
gratefully, in particular from CERN and the ATLAS Tier-1 facilities at
TRIUMF (Canada), NDGF (Denmark, Norway, Sweden), CC-IN2P3 (France),
KIT/GridKA (Germany), INFN-CNAF (Italy), NL-T1 (Netherlands), PIC (Spain),
ASGC (Taiwan), RAL (UK) and BNL (USA) and in the Tier-2 facilities
worldwide.

\printbibliography
\newpage
\begin{flushleft}
{\Large The ATLAS Collaboration}

\bigskip

M.~Aaboud$^\textrm{\scriptsize 136d}$,
G.~Aad$^\textrm{\scriptsize 87}$,
B.~Abbott$^\textrm{\scriptsize 114}$,
J.~Abdallah$^\textrm{\scriptsize 65}$,
O.~Abdinov$^\textrm{\scriptsize 12}$,
B.~Abeloos$^\textrm{\scriptsize 118}$,
R.~Aben$^\textrm{\scriptsize 108}$,
O.S.~AbouZeid$^\textrm{\scriptsize 138}$,
N.L.~Abraham$^\textrm{\scriptsize 150}$,
H.~Abramowicz$^\textrm{\scriptsize 154}$
% ...
\bigskip
\\
% ...
$^{12}$ Institute of Physics, Azerbaijan Academy of Sciences, Baku, Azerbaijan\\
% ...
$^{65}$ University of Iowa, Iowa City IA, United States of America\\
% ...
$^{87}$ CPPM, Aix-Marseille Universit{\'e} and CNRS/IN2P3, Marseille, France\\
% ...
$^{108}$ Nikhef National Institute for Subatomic Physics and University of Amsterdam, Amsterdam, Netherlands\\
% ...
$^{114}$ Homer L. Dodge Department of Physics and Astronomy, University of Oklahoma, Norman OK, United States of America\\
% ...
$^{118}$ LAL, Univ. Paris-Sud, CNRS/IN2P3, Universit{\'e} Paris-Saclay, Orsay, France\\
% ...
$^{136}$ $^{(a)}$ Facult{\'e} des Sciences Ain Chock, R{\'e}seau Universitaire de Physique des Hautes Energies - Universit{\'e} Hassan II, Casablanca; $^{(b)}$ Centre National de l'Energie des Sciences Techniques Nucleaires, Rabat; $^{(c)}$ Facult{\'e} des Sciences Semlalia, Universit{\'e} Cadi Ayyad, LPHEA-Marrakech; $^{(d)}$ Facult{\'e} des Sciences, Universit{\'e} Mohamed Premier and LPTPM, Oujda; $^{(e)}$ Facult{\'e} des sciences, Universit{\'e} Mohammed V, Rabat, Morocco\\
% ...
$^{138}$ Santa Cruz Institute for Particle Physics, University of California Santa Cruz, Santa Cruz CA, United States of America\\
% ...
$^{150}$ Department of Physics and Astronomy, University of Sussex, Brighton, United Kingdom\\
% ...
$^{154}$ Raymond and Beverly Sackler School of Physics and Astronomy, Tel Aviv University, Tel Aviv, Israel\\
% ...
$^{a}$ Also at Department of Physics, King's College London, London, United Kingdom\\
$^{b}$ Also at Institute of Physics, Azerbaijan Academy of Sciences, Baku, Azerbaijan\\
$^{c}$ Also at Novosibirsk State University, Novosibirsk, Russia\\
$^{d}$ Also at TRIUMF, Vancouver BC, Canada\\
$^{e}$ Also at Department of Physics {\&} Astronomy, University of Louisville, Louisville, KY, United States of America\\
% ...
\end{flushleft}


\end{document}